\pdfoutput=1
\documentclass[aps,prl,superscriptaddress,twocolumn,twoside,longbibliography]{revtex4-1}
\usepackage[utf8]{inputenc}

\usepackage{mathrsfs,mathtools,bbold}
\usepackage{amsmath,amssymb}
\usepackage{amsfonts}
\usepackage{verbatim}
\usepackage{float}
\usepackage{graphicx,epstopdf}
\usepackage{graphicx}
\usepackage{epsfig}

\usepackage{slashed}

\usepackage{color}

\newcommand{\ii}{\mathrm{i}}

\begin{document}

\title{How many surface modes does one see on the boundary of a Dirac material?}

\author{Maxim Kurkov}
\email{max.kurkov@gmail.com}

\affiliation{Dipartimento di Fisica ``Ettore Pancini'', Universit\`{a} di Napoli {\sl Federico II}, 
Napoli, Complesso Univ. Monte S. Angelo, Via Cintia, I-80126 Napoli, Italy}
\affiliation{INFN, Sezione di Napoli, Via Cintia, 80126 Napoli, Italy}

\author{Dmitri Vassilevich}
\email{dvassil@gmail.com}

\affiliation{CMCC, Universidade Federal do ABC, Santo Andr\'e, S\~ao Paulo, Brazil}
\affiliation{Physics Department, Tomsk State University, Tomsk, Russia}

\date{\today}

\begin{abstract}
We present full expressions for the surface part of polarization tensor of a Dirac fermion confined in a half-space in $3+1$ dimensions. We compare this tensor to the polarization tensor of eventual surface mode (which is a $2+1$ dimensional Dirac fermion) and find essential differences in the conductivities in both Hall and normal sectors. Thus, the interaction with electromagnetic field near the boundary differs significantly in the full model and in the effective theory for the surface mode.
\end{abstract}

\maketitle

\section{Introduction}
Modern interest to the physics of Dirac fermions occupying a bounded region in space is mostly related to new advanced materials, such as topological insulators (TIs) - see  reviews \cite{Hasan:2010xy,Qi:2011zya} and a monograph \cite{Bernevig:2013}. A lot of exciting physics of TIs is due to the presence of surface modes, that are also Dirac fermions though in one dimension less. Suppose that the Dirac fermion in a $3+1$ dimensional manifold has certain number of surface modes. Do we really see these modes by looking at the boundary? The interaction with photons is defined by the polarization tensor of fermions. Hence, we can rephrase the question as: What is the relation between the boundary part of the polarization tensor in $3+1$ dimensions and the polarization tensor of $2+1$ dimensional fermions? It is common to assume that the latter at least gives a good approximation to the former, see e.g. \cite{Li_2013,Li:2013wwg,Grushin:2015jja}.

The Dirac cones of surface modes of TIs are clearly seen with the spin-resolved ARPES methods \cite{Hsieh:2009,Xu_2011,Jozwiak_2011}. These cones are distorted as compared to the ideal case. However, such distortions can be taken into account by adding suitable correction terms to the Dirac Hamiltonian of surface modes \cite{Li_2013,Li:2013wwg}.

There is also another issue that has not received sufficient attention in the literature. There is no compelling reason to believe that the surface conductivity or the polarization tensor coincide with the conductivity or the polarization tensor computed in the effective $2+1$ dimensional theory. In principle, these two conductivities may be very different, so that one cannot be even considered as a reasonable approximation to the other. Thus, the study of relations between surface conductivities and effective conductivities of $2+1$ dimensional fields seems to be an important and timely problem.

The purpose of this work is to compare the polarization tensors (conductivities) in a model, which admits a complete analytic solution.
We take the model as in \cite{Fialkovsky:2019rum} with the {Lagrangian} $\mathcal{L}_{3+1}=\bar \psi \slashed{D}\psi$,
\begin{equation}
\slashed{D}=\ii \tilde\gamma^\mu(\partial_\mu +\ii e A_\mu) +im_5\gamma_5 +m.\label{Dirop}
\end{equation}
Here and in what follows tilde over a vector means rescaling of all spatial components with the Fermi velocity $v_F$, $\tilde\gamma^\mu \equiv \eta_\nu^\mu \gamma_\nu$, $\eta=\mathrm{diag}(1,v_F,v_F,v_F)$. As usual, $4\times 4$ $\gamma$ matrices satisfy $\{ \gamma^\mu,\gamma^\nu\}=2g^{\mu\nu}$ with $g=\mathrm{diag}(1,-1,-1,-1)$. $A_\mu$ is the electromagnetic potential. We shall mostly work in the units $c=1=\hbar$. Let us assume that the fermions propagate in a half-space $x^1>0$ satisfying the bag boundary conditions \cite{Chodos:1974je}
\begin{equation}
\tfrac 12 (1+\ii \gamma^1)\psi\vert_{x^1=0}=0.\label{bag}
\end{equation}
The bulk states have a mass gap $\mathfrak{m}=\sqrt{m^2+m_5^2}$. For $m<0$ there is also a boundary mode with the wave function decaying as $e^{m x^1/v_F}$ and satisfying the $(2+1)$ dimensional Dirac equation with respect to the coordinates $x^0$, $x^2$ and $x^3$ with the same Fermi velocity $v_F$ and with the mass given by $m_5$. It is important to realize that besides of giving a mass to the surface state $m_5$ is also essential for renormalization. The Pauli-Villars (PV) regulator fields must have axial masses to regularize the singularities in the polarization operator \cite{Fialkovsky:2019rum}. 

Note, that one may flip the sign in front {of} $\ii \gamma^1$ in (\ref{bag}) and obtain another set of admissible boundary conditions. As a result, one should change several signs in the effective action below. 

The interaction of electromagnetic field {with} the material is defined by the (one-loop) quantum effective active action of fermions, $S_{\rm eff}(A)=-\ii\, \mathrm{det}(\slashed{D}(A))$, which we truncate to the quadratic order in $A$ (since it is enough to describe most of the important physics). It is convenient to split the effective action as $S_{\rm eff}=S_{\rm bulk}+S_{\rm odd}+S_{\rm even}$. The first term
\begin{equation}
S_{\rm bulk}=\int d^4xd^4y F_{\mu\nu}(x) P^{\mu\nu\rho\sigma}(x,y)F_{\rho\sigma}(y)\label{Sbulk}
\end{equation}
depends on the kernel $P^{\mu\nu\rho\sigma}(x,y)$ which is exactly the same as in the theory without boundaries (though the integration in (\ref{Sbulk}) runs over the half-space). The terms $S_{\rm odd}$ and $S_{\rm even}$ appear due to the boundary. To obtain $S_{\rm odd}$ (respectively, $S_{\rm even}$) one has to collect the terms containing odd (respectively, even) number of $\gamma$-matrices in corresponding Feynman diagrams. As we shall see below, $S_{\rm odd}$ describes the Hall conductivity, while $S_{\rm even}$ describes the normal one. The corresponding polarization tensors are defined as usual,
\begin{equation}
S_{\rm odd/even}=\frac 12 \int d^4xd^4y A_\mu (x) \Pi^{\mu\nu}_{\rm odd/even}(x,y) A_\nu (y),\label{PT}
\end{equation}
{where the kernels $\Pi_{\rm odd/even}$ decay rapidly away from the boundary.}
We impose the axial gauge conditions $A_1=0$. The electronic properties of the boundary may be described by the boundary polarization tensor $\widehat\Pi$ which is obtained from $\Pi$ by integrating over the normal coordinates, 
\begin{equation}
\widehat \Pi^{jk}_{\rm odd/even}(x^i-y^i)=\int\limits_0^\infty dx^1\int\limits_0^\infty dy^1\, \Pi^{jk}_{\rm odd/even}(x,y),
\end{equation}
where $\{ i,j,k\}=\{ 0,2,3\}$.  We pass to the momentum representation of the polarization tensors
\begin{equation}\widehat\Pi^{jk}_{\rm odd/even}(p)=\int d^3 x\, e^{-\ii p_jx^j} \widehat \Pi^{jk}_{\rm odd/even}(x^i).
\end{equation}
This makes physical properties of these objects much more transparent.

In \cite{Fialkovsky:2019rum} the renormalized expressions for both $\Pi_{\rm odd}$ and $\Pi_{\rm even}$ were obtained in the coordinate representation. In the same representation an expression for the integrated kernel $\widehat\Pi_{\rm odd}$ was also presented. The main technical advance of the present article is the Fourier transformed expressions for both integrated kernels (see the form factors (\ref{Phi}), (\ref{Psirest}) and (\ref{Psim})). 

\section{Hall conductivity}
The $\widehat\Pi_{\rm odd}$ part of the polarization tensor can be expressed through a form factor $\Phi$, see eq.\ (\ref{Phi}), as
\begin{equation}
\widehat\Pi_{\rm odd}^{jl}(p)=\ii \alpha \varepsilon^{jlk}p_k \Phi(\omega),\label{Piodd}
\end{equation}
where $\alpha \equiv e^2/(4\pi)$ and $\omega=\sqrt{p_0^2-v_F^2(p_2^2+p_3^2)}$. The form factor reads
\begin{equation}
\Phi (\omega) =\frac{1}{2} \Phi_{3}(\omega)
-\frac{4 m_5}{\pi \omega} \int\limits_0^{m}d \beta \frac{\mathrm{arcsin}{\left(\frac{\omega}{2\sqrt{m_5^2+\beta^2}}\right)}}{\sqrt{-\omega^2 + 4(m_5^2+\beta^2)}}, \label{Phi}
\end{equation}
where 
\begin{equation}
\Phi_{3}(\omega) = \frac{2 m_{5}}{\omega}\,\mathrm{arctanh}{\left(\frac{\omega}{2|m_{5}|}\right)} - \mathrm{sgn}(m_{5})\label{Phi3}
\end{equation}
is the form factor for the polarization tensor in $2+1$ dimensions. This means, that the polarization tensor of a $2+1$ dimensional Dirac fermion can be obtained by replacing $\Phi$ by $\Phi_3$ in (\ref{Piodd}). 
The form factor (\ref{Phi}) has been obtained by making a Fourier transform in the coordinate space expression \cite[Eq.\ (74)]{Fialkovsky:2019rum}.

The expression (\ref{Phi}) is quite simple and may be evaluated at least numerically for all values of the parameters. The most interesting information is encoded in $\Phi(0)=-(1/\pi) \mathrm{arctan} (m/m_5)$. The corresponding surface Hall conductivity reads
\begin{equation}
\sigma_H=-\frac {e^2}{2\pi h}\mathrm{arctan} (m/m_5). \label{siH}
\end{equation}
(Here we restored the dependence on Planck constant $h$.) 
This result is fully consistent with the computations of boundary parity anomaly \cite{Kurkov:2017cdz} and with a singular limit of the fractional fermion charge of domain walls \cite{MateosGuilarte:2019eem}. In the case of a small surface mass gap,
$|m_5|\ll |m|$, we have $\sigma_H=-(e^2/4h) \mathrm{sgn}(m_5)\,\mathrm{sgn}(m)$. This is $\pm 1/2$ the value of Hall conductivity for a massless $2+1$ dimensional Dirac fermion. (The latter one is essentially defined by the parity anomaly \cite{Niemi:1983rq,Redlich:1983dv,AlvarezGaume:1984nf}.) This $1/2$, strange as it seems, nevertheless passes the following simple test \cite{Kurkov:2017cdz}. Consider a single Dirac fermion in a slab $0\leq x^1 \leq l$ in $3+1$ dimensions. This field can be represented as an infinite tower of $2+1$ dimensional modes with masses depending on $l$. In the limit $l\to 0$ all modes may become infinitely massive, or a one mode may remain massless, depending on the boundary conditions at two sides of the slab. Also depending on the boundary conditions, the Hall conductivities on opposite sides of the slab may either add up or compensate each other. One can show \cite{Kurkov:2017cdz} that the Hall conductivities add up precisely if there is a massless mode in the $l\to 0$ limit. The conductivities cancel against each other in the opposite case. We stress that these arguments rely on the fact that in the gapless case the Hall conductivity is defined through the parity anomaly which is local. Thus the conductivities on the boundaries do not depend on the distance between them.
 
The bulk time-reversal (TR) invariance is violated in the model considered here explicitly by the presence of $m_5$ and implicitly by axial masses of the PV regulator fields. Hence the restrictions on the boundary Hall conductivity derived in \cite{Qi:2008ew} {for} the theories with TR invariance in the bulk do not apply here even at the point $m_5=0$. Let us stress, that in the presence of boundaries the PV regulators must have axial masses \cite{Fialkovsky:2019rum}.

\section{Normal conductivity}
The ``even" part of polarization tensor reads
\begin{equation}
\widehat\Pi^{jk}_{\rm even}(p)=\frac {\alpha}{v_F^2} \eta^j_i\eta^k_l \left(
g^{il}-\frac{\tilde p^i\tilde p^l}{\tilde p^2} \right) \Psi (\omega).\label{Peven}
\end{equation}
It is convenient to subdivide the form factor as $\Psi=\Psi_{\rm mirr}+\Psi_{\rm rest}$ with
 \begin{eqnarray}
&&\!\!\!\!\!\!\!\!\!\!\!\!\Psi_{\mathrm{rest}}(\omega) = \frac{1}{2}\left(\Psi_{3}(\omega;m_{5})  -  \Psi_{3}(\omega;\mathfrak{m})\right) \nonumber  \\
&&\!\!\!\!\!\!\!\!\!\!\!\!-\frac{2}{ \pi} 
\int\limits_0^{m} d \beta \left(  1 - \frac{4(m_5^2+\beta^2)}{\omega}
\frac{\mathrm{arcsin}{\left(\frac{\omega}{2\sqrt{m_5^2+\beta^2}}\right)}}{\sqrt{-\omega^2 + 4(m_5^2+\beta^2)}}
  \right), \label{Psirest}
 \end{eqnarray}
 and
 \begin{equation} 
\Psi_{\rm mirr}(\omega)=-\frac {\omega^2}{2\pi} \int_0^{\frac \pi 2} d\xi \frac{\sin^4(\xi)}{\sqrt{4\mathfrak{m}^2-\omega^2\sin^2(\xi)}} \label{Psim},
\end{equation}
{where}
\begin{equation}
\Psi_{3}(\omega;m_5) = |m_5| 
- \frac{4m_5^2 + \omega^2 }{2\omega} \mathrm{arctanh}{\bigg(\frac{\omega}{2|m_5|} \bigg)}.\label{Psi3}
\end{equation}
To get (\ref{Psirest}) and (\ref{Psim}) we have integrated the polarization tensor from \cite{Fialkovsky:2019rum} over the normal coordinates and performed the Fourier transform in the tangential directions.  The parity-even part of polarization tensor for a Dirac fermion in $2+1$ dimensions is obtained if one substitutes $\Psi_3$ instead of $\Psi$ in (\ref{Peven}).

 The contribution corresponding $\Psi_{\rm mirr}$ is obtained by replacing the kernel $P^{\mu\nu\rho\sigma}(x,y)$ in (\ref{Sbulk}) (without renormalization terms) by its ``mirror image", see \cite{Fialkovsky:2019rum} for details. This form factor depends on the total bulk mass $\mathfrak{m}$ rather than on $m$ and $m_5$ separately. All other contributions are collected in $\Psi_{\rm rest}$. The characteristic feature of $\Psi_{\rm rest}$ is that it vanishes at $m=0$.

\begin{figure}[t]
\begin{minipage}[h]{0.49\linewidth}
\center{\includegraphics[width=1\linewidth]{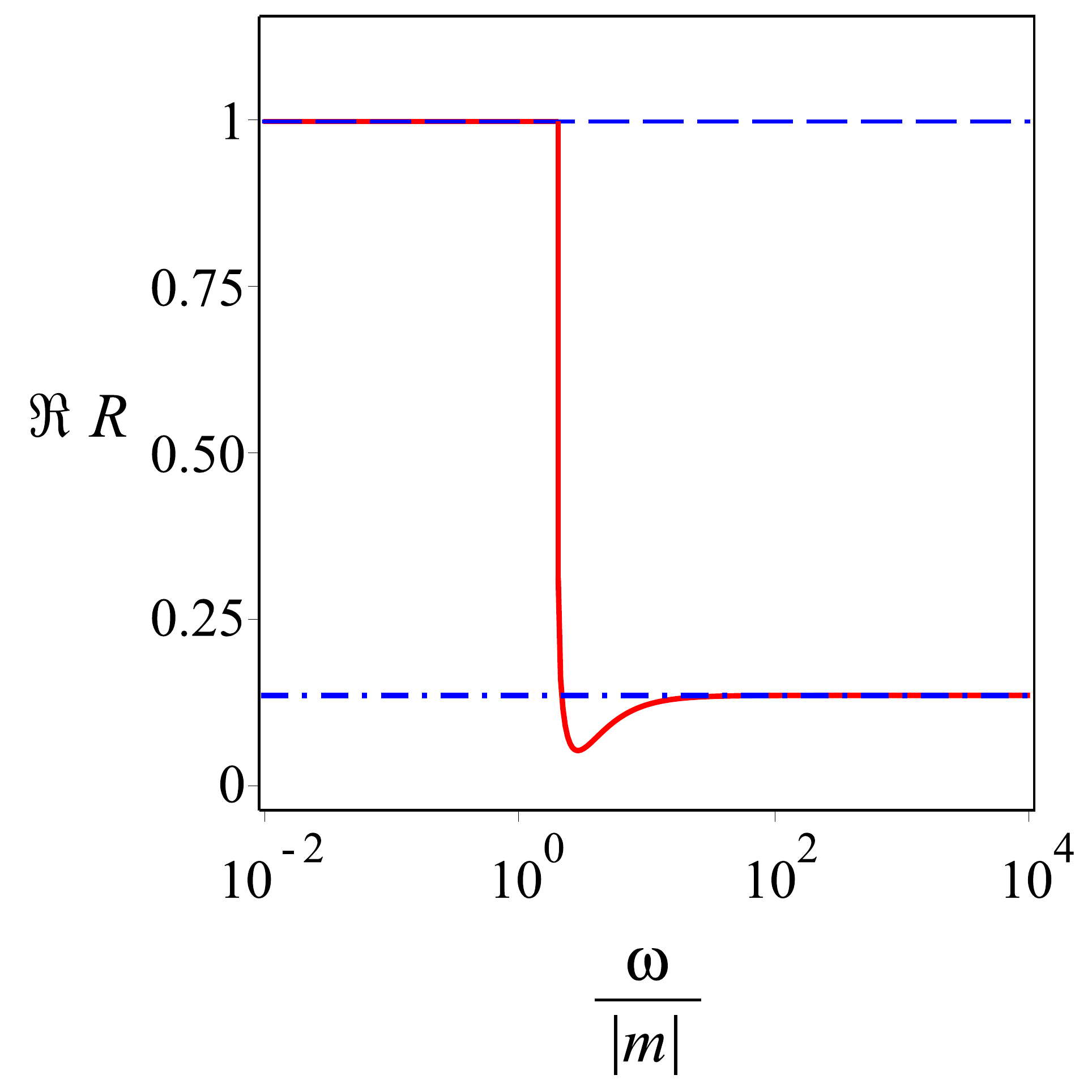}} a) \\
\end{minipage}
\hfill
\begin{minipage}[h]{0.49\linewidth}
\center{\includegraphics[width=1\linewidth]{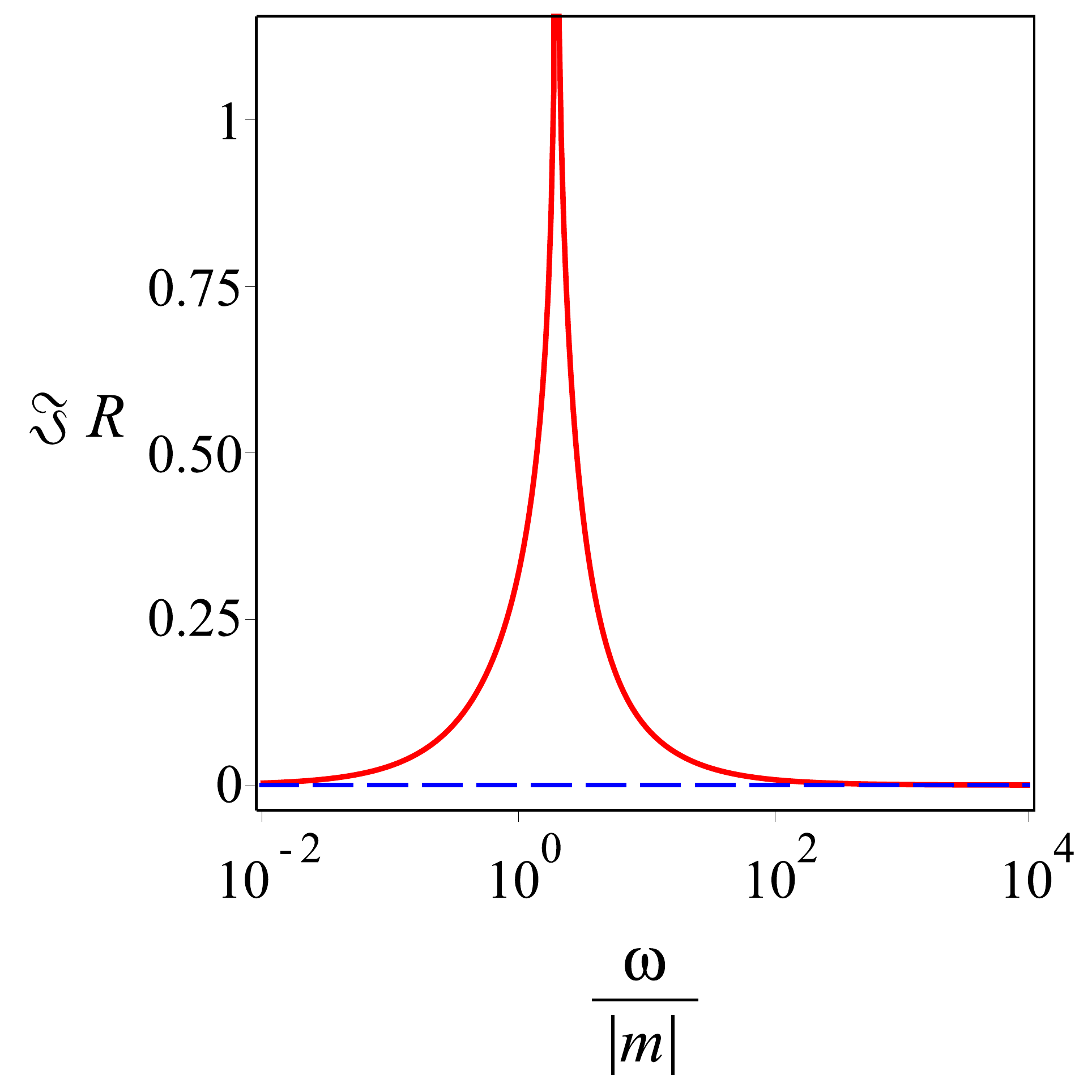}} \\b)
\end{minipage} \vspace{3mm}
\vfill
\begin{minipage}[h]{0.49\linewidth}
\center{\includegraphics[width=1\linewidth]{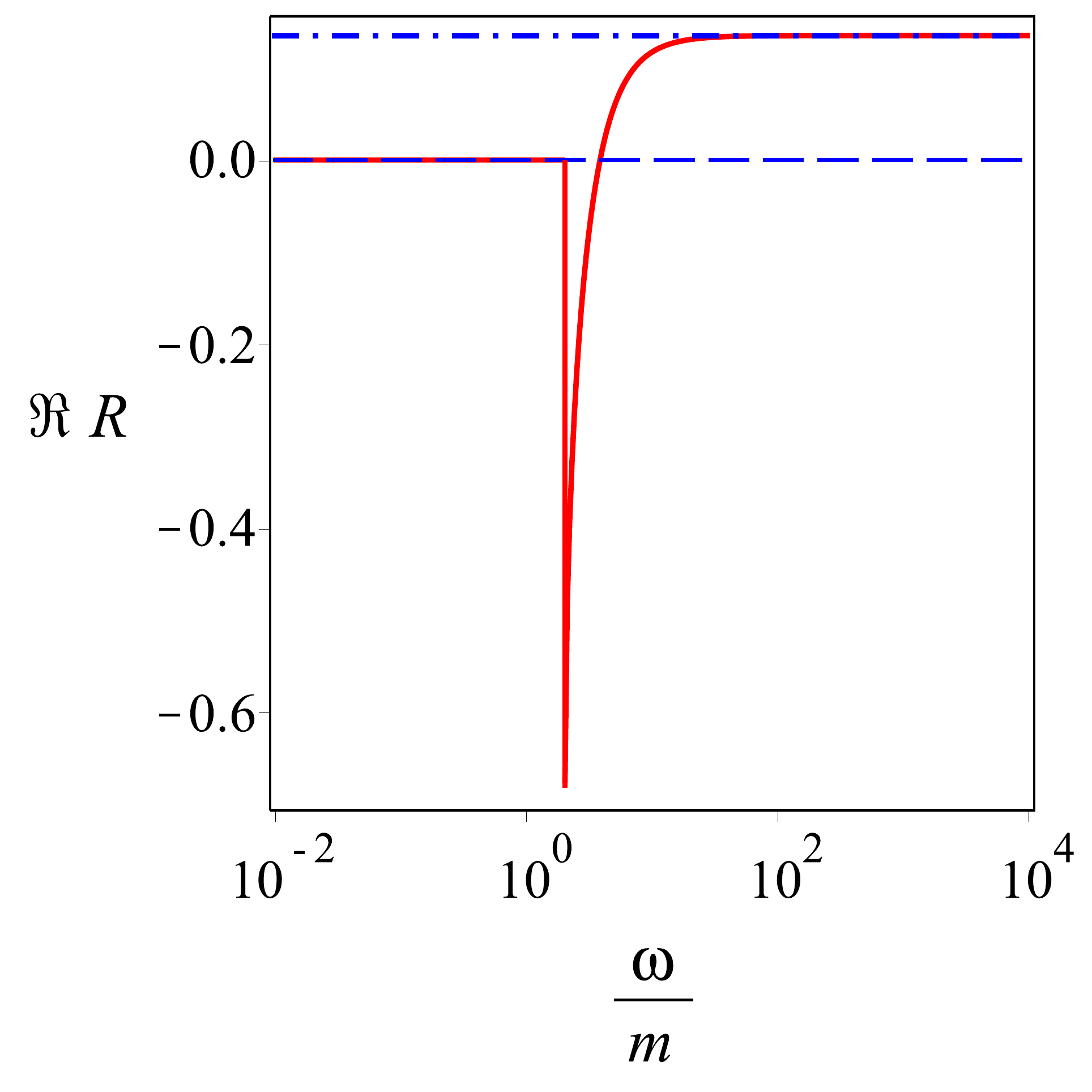}} c) \\
\end{minipage}
\hfill
\begin{minipage}[h]{0.49\linewidth}
\center{\includegraphics[width=1\linewidth]{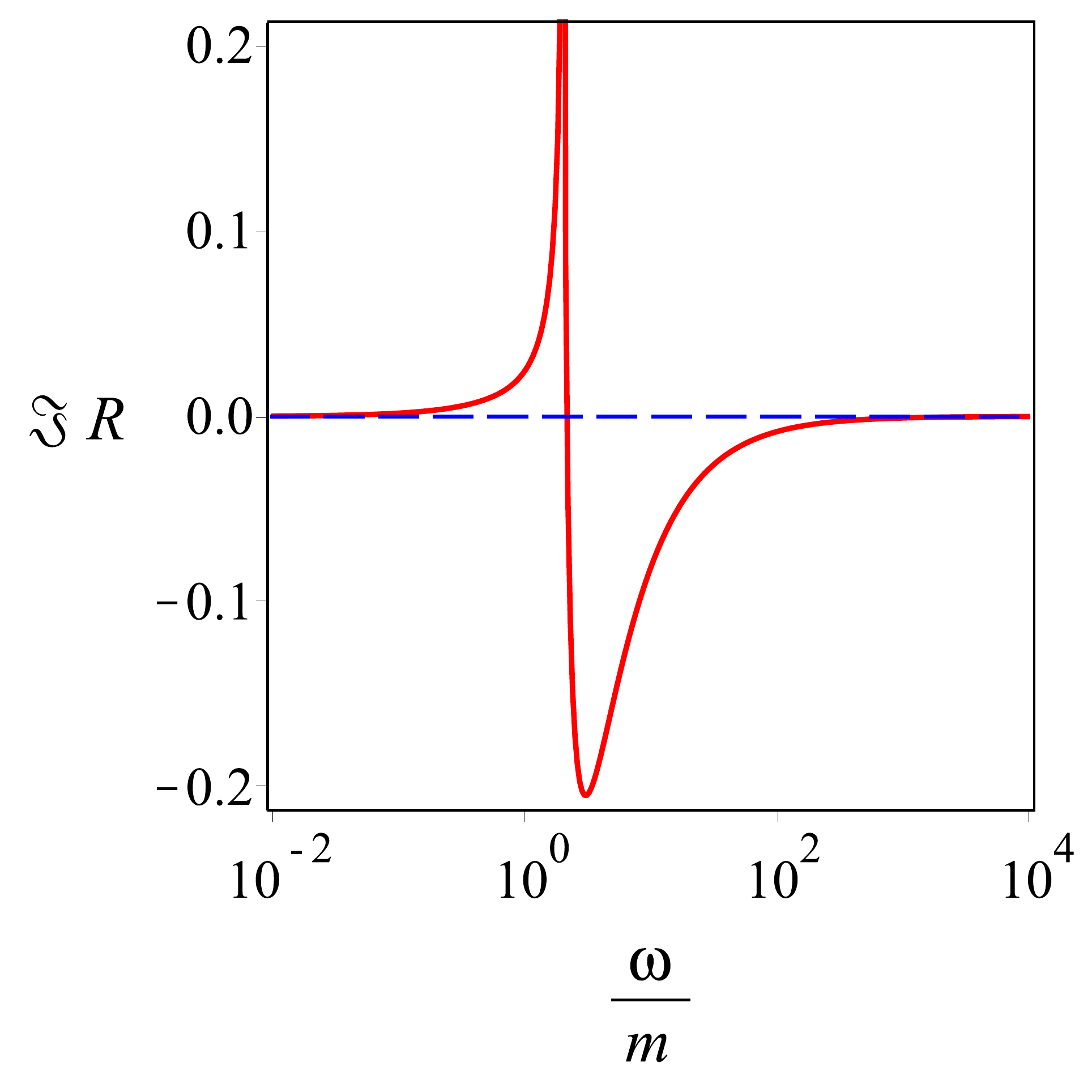}} d) \\
\end{minipage}
\caption{\sl Solid curves represent real and imaginary parts of the fraction $R(\omega)=\Psi_{\rm even}(\omega)/\Psi_3(\omega)$: 
a) $\Re{R}$ at $m<0$, b) $\Im{R}$ at $m<0$, c) $\Re{R}$ at $m>0$, d)$\Im{R}$ at $m>0$. Dash and dash-dot lines correspond to the infrared and ultraviolet limits respectively. 
 \label{pic}
}
\label{ris:experimentalcorrelationsignals}
\end{figure}

It is instructive to go to the case $m_5=0$ with gapless boundary modes and compare the form factors $\Psi$ and $\Psi_3$. For  $m_5=0$ one easily computes $\Psi_3(\omega)=-\ii \omega \pi/4$ (which corresponds to the universal DC conductivity of $2+1$ dimensional Dirac fermions $\sigma_{\rm DC}=e^2 \pi/(8h)$). As we see on Fig.~\ref{pic}, for small frequencies $\omega<2|m|$ the behaviour of the fraction
\begin{equation} 
 R(\omega)=\Psi_{\rm even}(\omega)/\Psi_3(\omega)\end{equation} 
 corresponds to the expectations. For negative $m$ (when there is a surface mode) the real part of surface conductivity corresponds to that of a single Dirac field in $2+1$ dimensions, while for $m>0$ (when there are no boundary modes) the real part of surface conductivity vanishes. In this region, $\Psi_{\rm rest}$ dominates over $\Psi_{\rm mirr}$. Near $\omega =2|m|$ there are singularities of $R$ -- a jump in the real part and a $\log$ singularity in the imaginary part. The structure of these singularities is typical for two point functions in quantum field theory at $\omega$ twice the mass gap. Here, this is the mass gap in the bulk rather than on the boundary. However, since $\Psi$ is constructed form the same blocks as the bulk polarization tensor (which is most obvious for the mirror contribution, but is also true for $\Psi_{\rm rest}$), the appearance of singularities at $2|m|$ is quite natural. At large frequencies, $\Psi_{\rm rest}$ becomes negligible. Thus, $R$ is defined by $\Psi_{\rm mirr}$ and does not depend on the sign on $m$. The asymptotic values are $\Re R=4/(3\pi^2)$ and $\Im R=0$. 

For a non-zero but small $m_5$ the behaviour of $\Re R(\omega)$ remains qualitatively the same. Although both $\Psi_{\rm rest}$ and $\Psi_3$ develop singularities at $\omega=2|m_5|$, the real part $\Re R(\omega)$ remains near $1$ for $m<0$ (respectively, near $0$ for $m>0$) and reasonably small frequencies.


\section{Conclusions}
The main technical result of this work is the expressions (\ref{Phi}), (\ref{Psirest}) and (\ref{Psim}) for the boundary form factors. We did not report here any details of the computations (which we hope to do somewhere else), but instead we concentrated on discussing the properties of surface conductivities.

Let us summarize our findings on the comparison of effective surface conductivities with conductivities of $2+1$ dimensional fermions. The boundary Hall conductivity (\ref{siH}) exhibits a strong dependence on the bulk mass gap which cannot occur in the effective $2+1$ dimensional theory. For a small surface gap $|m_5|\ll |m|$, the Hall conductivity depends on $m$ only through $\mathrm{sgn}(m)$, but has quarter-integer values instead of more common half-integer ones. One can derive, however, some interesting ``sum rules". For example, if $m$ changes sign from $m<0$ (one boundary mode) to $m>0$ (no boundary modes) still satisfying $|m|\gg |m_5|$, the zero-frequency Hall conductivity changes by the Hall conductivity of single massless $2+1$ dimensional mode.

The parity-even part of boundary polarization tensor at small frequencies, $\omega<2|m|$, reproduces the polarization tensor for right number of $2+1$ dimensional surface states. Thus we can say that in the even sector and for moderate frequencies one indeed sees the right number of surface modes. However, around $\omega\sim 2|m|$ the form factor develops singularities that are absent in the effective boundary theory. For $\omega\gg 2|m|$ this form factor goes to a universal asymptotics which does not depend on whether the theory has or has not a surface state for this particular sign of $m$. 

TIs and other Dirac materials usually contain in the continuous limit something more than just a single Dirac fermion. Thus, what we shall see will be a combined effect of several polarization tensors (as described here) with various parameters. {A famous example of this effect is graphene \cite{Semenoff:1984dq}.} However, even in combined expressions one shall see considerable deviations from the naive picture based on the conductivity of surface states. 
On a positive side, our results provide explicit and simple formulas for the surface conductivity. Thus abandoning the assumption that surface conductivity is given by conductivities of the $2+1$ dimensional modes does not lead to considerable technical complications.

The surface conductance of TIs is hard to measure due to first of all the residual bulk carrier density. However, the new advanced techniques \cite{cai2018independence,Seifert_2019} make the prospects of precision measurement of the surface conductance together with experimental verification of our results quite realistic. We also expect that optical experiments will reveal the resonance behaviour of longitudinal conductivity at the frequencies $\sim |m|$. 

Another place where our results will be useful is the computations of Casimir interaction between Dirac materials (see recent papers \cite{Fialkovsky:2018fpo,Marachevsky:2018axn,Rodriguez-Lopez:2019oex,Farias:2020qqp} and references therein), which are motivated by the exciting possibility to get a repulsive Casimir force. Note, that since the Casimir force is given by a momentum integral, it is important to have expressions for the polarization tensor at all frequencies, as we gave here.
\begin{acknowledgments}
We are grateful to Ignat Fialkovsky, Juan Mateos Guilarte and Nail Khusnutdinov for discussions and collaboration on related papers. The work of D.V.V. was supported in parts by the S\~ao Paulo Research Foundation (FAPESP), projects 2016/03319-6 and 2017/50294-1, by the grant 305594/2019-2 of CNPq, by the RFBR project 18-02-00149-a and by the Tomsk State University Competitiveness Improvement Program. M. K. acknowledges the support of the INFN Iniziativa Specifica GeoSymQFT.
\end{acknowledgments}


\bibliography{graphene}

\end{document}